\documentclass[aps,prd,twocolumn,superscriptaddress,amsmath,showpacs,floatfix,nofootinbib]{revtex4-1}
\usepackage{graphicx}
\usepackage{dcolumn}
\usepackage{bm}
\usepackage{natbib}
\usepackage{multirow}

\newcommand{\Mpc}{{\rm ~Mpc}}

\begin{document}

\title{Updating constraints on inflationary features in the primordial power spectrum
with the Planck data.}

\author{Micol Benetti}
\affiliation{Physics Department and ICRA, Universit\`a di Roma ``La Sapienza'', Ple.\ Aldo Moro 2, 00185, Rome, Italy}
\affiliation{Physics Department and INFN, Universit\`a di Roma ``La Sapienza'', Ple Aldo Moro 2, 00185, Rome, Italy}

\begin{abstract}
We present new constraints on possible features in the primordial inflationary density perturbations power spectrum in light of the  recent Cosmic Microwave Background Anisotropies measurements from the Planck satellite.  We found that the Planck data hints for the presence of features
in two different ranges of angular scales, corresponding to multipoles $10< \ell < 60$ and $150< \ell < 300$,
with a decrease in the best fit $\chi^2$ value with respect to the featureless "vanilla" $\Lambda$CDM model 
of $\Delta\chi^2 \simeq 9$ in both cases.
\end{abstract}
\maketitle

\section{Introduction} \label {sec:intro}

The recent results from the Planck satellite on the cosmic microwave background (CMB, hereafter)
angular power spectrum are in very good agreement with the theoretical expectations of the
simplest inflationary model based on a single, minimally coupled, scalar field \cite{Ade:2013zuv}.

However, as already discussed in \cite{Ade:2013uln}, some interesting hints for deviations from scale invariance are present in the Planck data and are certainly worthwhile of further investigation.

In this brief paper, we present new constraints on an inflationary model with step-like features 
as proposed by \cite{Adams:1996yd, Adams:1997de}. 

Step-like features in the inflationary potential are expected in theories with multiple interacting 
scalar fields as supergravity-inspired models, 
where supersymmetry-breaking phase transitions occur during inflation. At the same time,
step-like features are able to produce localized oscillations in the CMB angular power spectra
and, in particular, as we have already shown in \cite{Benetti:2012wu}, to provide
a better fit with respect to the featureless case in case of the WMAP data.
It is therefore timely to analyze the new Planck data, that covers a larger multipole range
respect to WMAP, and to quantify the compatibility of the features with this new dataset.

A first analysis has already been provided by the Planck collaboration in \cite{Ade:2013uln}.
However, as we will discuss in the next section, this analysis assumed an analytical and, therefore,
approximate formula for the features and investigated a range of angular scales different from the
one analyzed in \cite{Benetti:2012wu}. In particular, as we will discuss below, the
analysis presented in \cite{Ade:2013uln} for step-like feature did not cover the range of low
multipoles. Moreover, the remaining cosmological parameters
were not let to vary freely but fixed at their best fit values, therefore neglecting
possible correlations.

Here, on the contrary, we assume the same parameter range of \cite{Benetti:2012wu} and
we integrate the set of differential equations to accurately compute the
oscillations in the CMB angular spectrum, given a step-like feature in the 
inflationary potential (again, see \cite{Benetti:2012wu}). Moreover, we let 
all the parameters to vary freely,  taking into account possible correlations
between the parameters. For comparison, we also use the analytical model adopted in \cite{Ade:2013uln}.

The paper is organized ad follows: in Sec.II we briefly explain 
the analysis method adopted; in Sec.III we present the results of our analysis and in Sec IV 
we summarize our conclusions.\\
%
\begin{table*}
\centering
\caption{Best Fit values and $68\%$ confidence limits for the cosmological and step parameters. The second and third columns  refer to $\Lambda$CDM model; the fourth, fifth and sixth column show the constraints on the features model using the same prior ranges of \cite{Benetti:2012wu} with, respectively,  the numerical integration approach and the analytical approximate approach; the last two columns show the constraints obtained using the analytical approach and the parameter range of \cite{Ade:2013uln}. \label{tab:Tabel_one}}
\begin{tabular}{c|cc|cc|c|cc}
\hline
\hline

\multicolumn{1}{c|}{$$}&
\multicolumn{2}{c|}{\textbf{$\Lambda$CDM model}}& 
\multicolumn{2}{c}{\textbf{Numerical integration} \footnotemark[1]  }& 
\multicolumn{1}{c|}{\textbf{Approx} \footnotemark[2] \footnotemark[4]}&
\multicolumn{2}{c}{\textbf{Approximate parameterization} \footnotemark[3] \footnotemark[4] }\\ 
\textbf{Parameter}		& Best fit\footnotemark[5] &$68\%$ limits 	& Best fit\footnotemark[5] &$68\%$ limits 	& Best fit\footnotemark[5] 	& Best fit\footnotemark[5] & $68\%$ limits \\
\hline
$100\,\Omega_b h^2$ 	& $2.210$& $2.220\pm0.028$	& $2.220$& $2.220\pm0.028$	&$2.199$ 	&$2.220$ & $2.220\pm0.029$\\
$\Omega_{c} h^2$		& $0.1203$& $0.1199\pm0.0027$	& $0.1212$ & $0.1203\pm0.0028$ 	& $0.1212$ 	 & $0.1209$ & $0.1200\pm0.0026$\\
$100\, \theta$                         & $1.0413$& $1.0413\pm0.0006$	& $1.0411$& $1.0413\pm0.0006$ 	& $1.0410$ 	& $1.0410$& $1.0413\pm0.0006$\\
$\tau$ 		                  	& $0.090$& $0.090\pm0.013$	& $0.089$& $0.091\pm0.014$	& $0.092$	& $0.094$& $0.089\pm0.010$\\
$n_s$			  	& $0.963$& $0.961\pm0.007$	& $0.959$& $0.959\pm0.008$	& $0.958$ 	& $0.960$ & $0.960\pm0.007$ \\
$10^9 A_s$\footnotemark[6] & $2.21$& $ 2.20\pm0.05$		& $ 2.20$& $2.21\pm0.06$  		& $ 2.22$	& $ 2.22$& $2.22\pm0.04$ \\
$b$ 		               & $ - $& $ - $  				&$14.66$& $14.99\pm0.29$		& $ - $		& $ - $& $ - $\\
$\log c$ 		  & $ - $& $ - $   				&$-2.85$& $-2.99\pm0.61$		& $ - $		& $ - $	& $ - $ \\	
$\log d$ 		   & $ - $ & $ - $  				&$-1.44$& $-1.32\pm0.54$		& $ - $		& $ - $ & $ - $\\
$A_f$ 		               & $ - $& $ - $  				& $ - $& $ - $ 				& $0.90$	& $0.10$& $0.10\pm0.06$\\
ln $\eta_f / Mpc$ 	 & $ - $& $ - $  				& $ - $ & $ - $ 				& $7.17$	& $7.25$& $6.34\pm3.7$  \\
ln $x_d$ 		    & $ - $& $ - $  				& $ - $& $ - $				& $0.4$	& $4.47$& $1.86\pm1.74$\\
Age [Gyr] 		   & $13.82$ & $13.82\pm0.05$		& $13.83$ & $13.82\pm0.05$ 	& $13.83$ 	& $13.84$& $13.82\pm0.05$ \\
$z_{re}$ 		& $11.3$ & $11.1\pm1.1$			& $11.1$& $11.2 \pm1.2$ 		& $11.4$	& $11.5$& $11.1 \pm0.9$\\
$H_0 $ [km s$^{-1}$ Mpc$^{-1}$] 	 & $67.2$& $67.3\pm1.2$ 	& $66.7$& $67.1\pm1.2$ 		& $66.7$ 	& $66.7$& $67.3\pm1.2$ \\
\\
\hline
$- 2\log\mathcal{L}$         & $9803$ &$$	& $9794$  & $$ & $9794$  	& $9793$  &$$\\
\hline
\hline
\footnotetext[1]{Uses initial potential as Eq.(1).}
\footnotetext[2]{Uses low-$\ell$ priors.} 
\footnotetext[3]{Uses Planck \cite{Ade:2013uln} priors.}
\footnotetext[4]{Uses Approximate parameterization as Eq(2).}
\footnotetext[5]{Calculated using BOBYQA algorithm.}
\footnotetext[6]{$k_0 = 0.05\,\Mpc^{-1}$.}
\end{tabular}
\end{table*}

\section{Model and Analysis method} \label {sec:Model and Analysis method}
 
Following the work of Adams \emph{et al.} \cite{Adams:2001vc}, we consider a model with a step-like 
feature added to a chaotic potential $V(\phi)=m^2\phi^2/2$, for the inflaton field $\phi$, of the form:

\begin{equation}
V(\phi) = \frac{1}{2}m^2\phi^2 \left[1+ c\tanh\left(\frac{\phi-b}{d}\right)\right] \, ,
\label{eq:Vstep}
\end{equation}

where $b$ is the value of the field where the step is located, $c$ is the height of the step and $d$ its slope. 

In order to evaluate the density perturbation spectrum we numerically evolve the 
relevant equations that, for brevity, we do not report here and we refer the reader to
\cite{Benetti:2012wu, Leach:2000yw, Leach:2001zf, Adams:2001vc,peiris,hamann, covi,Mortonson:2009qv,Hazra:2010ve,Ashoorioon:2006wc}.

Moreover, we also adopt an analytical parameterization for the scalar primordial power spectrum 
given by \cite{Adshead:2012xz, Adshead:2011jq}:

\begin{subequations}

\begin{equation}
P_R(k) = \exp [ \ln P_0 (k) + \frac{A_f}{3} \frac {k\eta_f}{\sinh\ (\frac{k\eta_f}{x_d})} W'(k\eta_f)]  , 
\label{eq:Adshead1} \\
\end{equation}
\begin{equation}
W'(x) = (-3+ \frac {9}{x^2})\cos 2x + (15-\frac{9}{x^2}) \frac{\sin 2x}{2x} 
\label{eq:Adshead2}
\end{equation}
\end{subequations}

where $P_0 (k) = A_s (\frac{k}{k_*})^ {n_s - 1}$ is the smooth spectrum with the standard power law form, $A_f$ is the kinetic energy perturbation of the step, $\eta_f$ is the step crossing time in units of Mpc and $x_d$ the dimensionless damping scale. 
Using this method, by placing the features directly on the density power spectrum, we do not integrate 
the system of differential equations, with a significantly smaller computing time. This is the same approach that has been used in \cite{Ade:2013uln}.\\

We therefore consider a "vanilla" \newcommand{\LCDM}{\Lambda\mathrm{CDM}} theoretical model with the addition of  features in the primordial spectrum, parametrized in both cases (numerical and analytical) by three parameters.
Together with these parameter we vary the usual cosmological parameters as the baryon density, $\omega_b$, 
the cold dark matter density, $\omega_c$, the ratio between the sound horizon and the angular diameter distance at decoupling, $\theta$, 
the optical depth, $\tau$, the primordial scalar amplitude, $\mathcal A_s$, and, finally the primordial spectral index $n_s$.
 We also vary the nuisance foregrounds parameters \cite{Planck:2013kta}, we consider purely adiabatic initial conditions,  fix the sum of neutrino masses to $0.06$ eV, and we limit the analysis to scalar perturbations.\\

We then perform a Monte Carlo Markov Chain analysis via the publicly available package \verb+CosmoMC+ \cite{Lewis:2002ah}. We use a modified version of the \verb+CAMB+ (\cite{camb}) code, needed to compute the CMB anisotropies spectrum for given values of the parameters describing this type of inflationary model. The Gelman and Rubin criteria is used to evaluate the convergence of the chains, demanding that $R -1 \leq 0.02$. 
By default CosmoMC uses a simple Metropolis-Hastings algorithm that needs to evaluate the model likelihood at each point traversed by the chains. It is designed to draw samples from the posterior distribution and not to find the best-fit model. Thus in the analysis we use the Bound Optimization BY Quadratic Approximation (BOBYQA) algorithm, developed by Powell \cite{Powell}, that is a optimized methods for minimizing functions of more variables and is implemented in CosmoMC. The quoted results in this paper for the best-fit values of the parameters, as well as for the value of the $\chi^2$ itself, are obtained using Powell' s routines. \\

The dataset considered in this work, available from the ESA website \footnotemark[1], are: \\
$-$ high-l Planck temperature ($50< \ell <2500$, derived from the CamSpec likelihood by combining spectra in the frequency range $100 - 217$ GHz \cite{Planck:2013kta}), \\
$-$ low-$\ell$ Planck temperature ($2< \ell <49$, derived from a component-separation algorithm, Commander, applied to maps in the frequency range $30 - 353$ GHz \cite{Ade:2013hta}),\\
$-$ low-$\ell$ WMAP-9year polarization \cite{wmap9}.\\
The likelihood code is provided by the Planck collaboration \cite{Planck:2013kta}. \\
\footnotetext[1]{\text{www.sciops.esa.int}}

The pivot wave-number selected is $k_\star=k_0 =0.05\,\mathrm{Mpc}^{-1}$ , which is the same value chosen by the Planck collaboration for this type of study. This parameter is degenerate with the value of the position of the step in $\phi$, e.g. changing $k_0$ from $0.05$ to $0.002$ Mpc$^{-1}$ shifts the step value $b$ by $\sim0.5$ towards lower values (see \cite{Benetti:2012wu}).

\begin{figure}
	\centering
	\includegraphics[width=1.1\hsize]{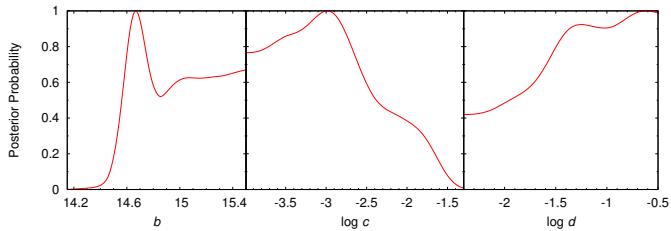}
	\caption{One-dimensional posterior probability densities for the step parameters of an inflationary model with step-like features in the potential, obtained by numerical integration of the mode equations.}
	\label{fig:features_posterior}
\end{figure}
\section{Results and discussion} \label {sec:Results and discussion}

We essentially consider three types of analysis with the results reported in Table 1.
The first analysis assumes a simple $\Lambda$CDM model with a featureless spectrum.
For the second analysis we considered the step-like model in the inflationary
potential, numerically integrating the relevant equations and assuming the
following priors on the corresponding parameters: 
$14.2 \leq b \leq 15.5$, $-4 \leq \log c \leq -1$, $-2.5 \leq \log d \leq -0.5$.
These results are reported in fourth and fifth columns of Table \ref{tab:Tabel_one}.
The comparison between the results presented in the two tables are useful
in order to identify the impact of primordial features on the constraints on the
standard $\Lambda$CDM parameters.

In the third analysis we used the analytical formula presented in 
\cite{Ade:2013uln} with the same choice of priors and given by: 
$0 \leq A_f \leq 0.2$ , $0 \leq $ ln($\eta_f/\mathrm{Mpc} ) \leq 12$, $-1 \leq {\ln}x_d  \leq 5$.
These values are reported in the last two columns of Table \ref{tab:Tabel_one}.

As we can see, introducing oscillations in the primordial spectrum either by numerical integration the
relevant equation or by using the above mentioned analytical formula, reduces the
$\chi^2$ of the best fit model by $\Delta \chi^2 \sim 9$. 
However, the feature parameters are poorly constrained, as also shown in Fig.\ref{fig:features_posterior}
where we report the posterior probabilities for the numerical integrating analysis.
Moreover,  the introduction of features has little effect on the constraints on the remaining, nuisance, cosmological parameters. 

In Fig.\ref{fig:features_posterior} we can also note that the posteriors are better defined respect to those present in our previous work of \cite{Benetti:2012wu} although they are significantly different from a gaussian distribution. In particular, we see that  the use of the Planck data eliminates a bimodal form in the posterior probability for the $b$ parameter,  present in the WMAP9 data.
\begin{figure}
	\centering
	\includegraphics[width=1\hsize]{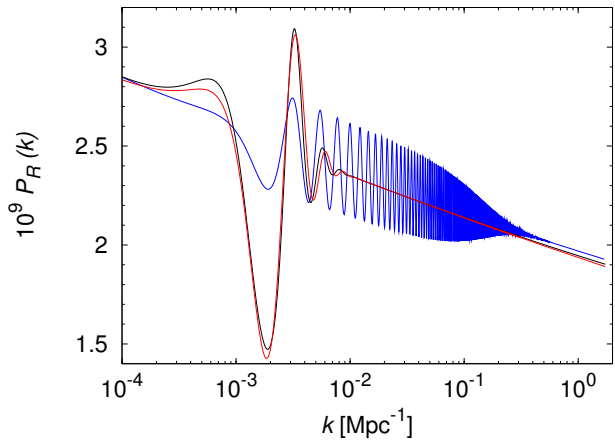}
	\caption{ Primordial power spectra for inflationary potentials with step. 
The \textbf{red line} shows the best-fit spectrum in the case of the numerical integration approach with $m = 7.5$x$10^{-6}$, using the prior ranges of \cite{Benetti:2012wu}, while the \textbf{black line} shows the best-fit spectrum of the respective approximate parameterization analysis. The \textbf{blue line} plots the best-fit approximate scalar power in the range of parameters of \cite{Ade:2013uln}, which causes features in $150< \ell <300$.}
	\label{fig:Pk}
\end{figure}
As we see in Table I, both the analytical and the numerical method provide the same reduction
in the $\chi^2$ value. In particular, the results for the analytical method, are fully consistent
with those reported in \cite{Ade:2013uln}.

However the effects on the CMB angular spectra are drastically different. The best-fit model
obtained from a numerical integration provide significantly different oscillations respect to
the best fit model obtained in the case of the analytical approximation.

We can clearly see this in Fig. \ref{fig:Pk}, where we plot the primordial power spectra for the
best fit models obtained in the case of numerical integration (red line) and for the case
of analytical approximation (blue line) used in the Planck analysis.

\begin{figure}
	\resizebox{0.6\textwidth}{!}{\input{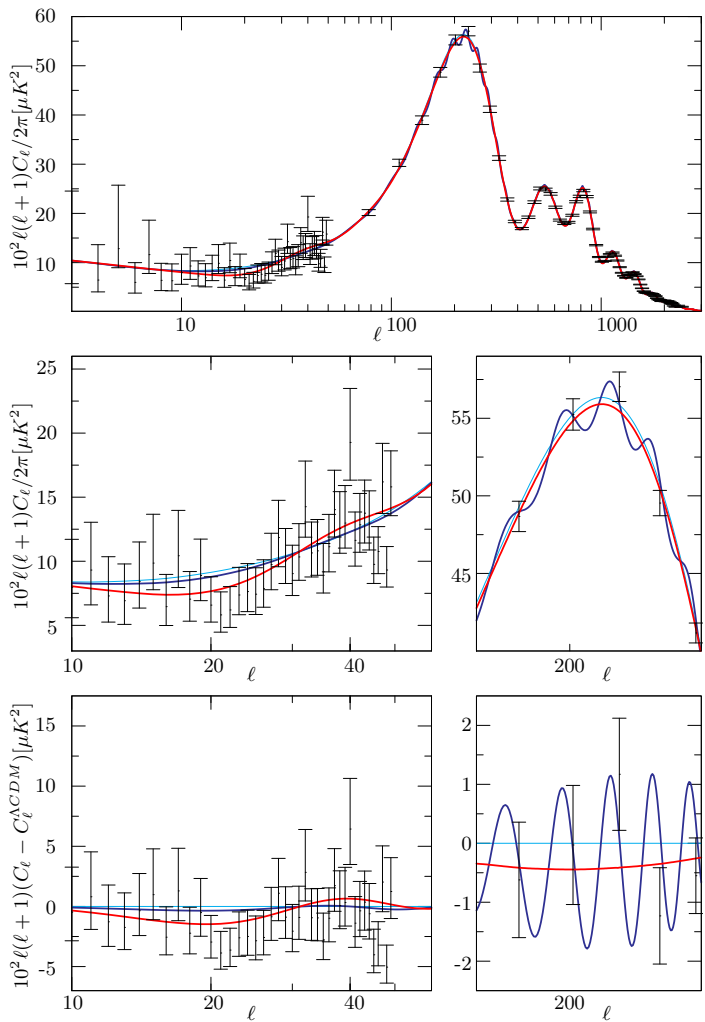}}
	\caption{Upper panel: Temperature power spectrum for the best-fit $\Lambda$CDM model (\textbf{light blue line}) and for two step models with features at low-$\ell$ (\textbf{red line}) and features at the first peak (\textbf{blue line}), corresponding to the best fit of the numerical integration (with the same priors as \cite{Benetti:2012wu}) and analytical approach (with the same priors as \cite{Ade:2013uln}), compared to the Planck temperature data. 
\\ Middle panel:  The same as above, zoomed in the region  $10< \ell <60$ and $150< \ell <300$.
\\ Lower panel: The same as above, plotted in terms of residuals with respect to the $\Lambda$CDM best-fit.}
	\label{fig:Powerspectrum}
\end{figure}

In Fig. \ref{fig:Powerspectrum} we compare the best fit 
CMB angular spectra obtained in the two cases. As we can see, the numerical integration
method identifies the oscillations on large angular scales ($10 < \ell < 60$) while
the analytical method provides a better fit by producing oscillations around the
first doppler peak.

This difference is essentially due to the different choice of priors on the feature parameters 
assumed in the two analyses. To check this, we changed the priors
for the analysis based on the analytical formula to
$0.8 \leq A_f \leq 1$ , $7 \leq $ ln($\eta_f/\mathrm{Mpc} )  \leq 8$, $0 \leq {\ln}x_d \leq 0.5$,
obtaining the best fit values reported in Table I, sixth column.
As we can see, the best fit has now $A_f\sim0.9$, a value that was excluded by the
choice of priors used in \cite{Ade:2013uln}.
The corresponding primordial spectrum is reported in Fig.\ref{fig:Pk}
as a black line and, as we can see, is in full agreement with the best fit spectra 
obtained from the analysis made assuming the numerical integration method.

We can therefore conclude that one needs to be extremely cautious in 
the choice of priors when looking for features in the CMB spectra since
 probability distributions for the parameters are highly multimodal.

\section{Concluding remarks} \label {sec:Concluding remarks}

We have presented updated constraints on an inflationary model with a step-like feature in the inflaton potential, using WMAP$9$ low-$\ell$ polarization data and the recent temperature data released by the Planck experiment. Such a feature would induce oscillations in the anisotropy power spectrum with magnitude, extent and position depending on three step parameters. 

We have considered two different methods.  The first uses a numerical routine to accurately calculate the primordial density spectrum corresponding to a given inflaton potential. The second employs an approximate form of the power spectrum, reproducing the features caused by a step-like inflaton potential step-like. For the latter analysis, we have also studied the impact of different prior ranges, corresponding to features in the low-$\ell$ and mid-$\ell$ ranges.

The analysis done performing the exact integration of the mode equations shows a minimum $\chi^2$ value with $\Delta\chi^2 \simeq 9$ with respect to the featureless $\Lambda$CDM model, at the cost of three new parameters. This improvement is due to the presence of oscillations in the multipole range $10< \ell <60$. These results can be matched using instead the analytical approach, by chosing a suitable prior range for the parameters, different from the one used in \cite{Ade:2013uln}. On the other hand, the results for the analytical method with the same prior range as \cite{Ade:2013uln}, corresponding instead to oscillations in the range $150< \ell <300$, are fully consistent with those reported there. The improvement in the goodness-of-fit is still $\Delta\chi^2 \simeq 9$, although it is caused by oscillations in a completely different range of scales.

Finally, the constraints on the step parameters are improved with respect to our previous work \cite{Benetti:2012wu}.

Future polarization data, as discussed in \cite{Benetti:2012wu} will clearly further
improve the constraints presented here.

Our results are in reasonable agreement, given the different method of analysis adopted,
also with those recently presented in \cite{meerburg}.

\section{Acknowledgments}
It is a pleasure to thank Jan Hamann for providing the numerical code that computes the primordial inflationary spectra. I would like to thank Carlo Luciano Bianco for useful discussion.



\begin{thebibliography}{99}

\bibitem{Ade:2013zuv} 
  P.~A.~R.~Ade {\it et al.}  [Planck Collaboration],
  arXiv:1303.5076 [astro-ph.CO].

\bibitem{Ade:2013uln}
  P.~A.~R.~Ade {\it et al.}  [Planck Collaboration],
  arXiv:1303.5082 [astro-ph.CO].

\bibitem{Adams:1996yd} 
  J.~A.~Adams, G.~G.~Ross and S.~Sarkar,
  Phys.\ Lett.\ B {\bf 391}, 271 (1997)
  [hep-ph/9608336].
 %
\bibitem{Adams:1997de} 
  J.~A.~Adams, G.~G.~Ross and S.~Sarkar,
  Nucl.\ Phys.\ B {\bf 503}, 405 (1997)
  [hep-ph/9704286].
%

\bibitem{Benetti:2012wu}
  M.~Benetti, S.~Pandolfi, M.~Lattanzi, M.~Martinelli and A.~Melchiorri,
  Phys.\ Rev.\ D {\bf 87} (2013) 023519
  [arXiv:1210.3562 [astro-ph.CO]].

\bibitem{Adams:2001vc} 
  J.~A.~Adams, B.~Cresswell and R.~Easther,
  Phys.\ Rev.\ D {\bf 64}, 123514 (2001)
  [astro-ph/0102236].
%
\bibitem{Leach:2000yw} 
  S.~M.~Leach and A.~R.~Liddle,
  Phys.\ Rev.\ D {\bf 63}, 043508 (2001)
  [astro-ph/0010082].

\bibitem{Leach:2001zf} 
  S.~MLeach, M.~Sasaki, D.~Wands and A.~RLiddle,
  Phys.\ Rev.\ D {\bf 64}, 023512 (2001)
  [astro-ph/0101406].

\bibitem{peiris}
  H.~V.~Peiris {\it et al.}  [WMAP Collaboration],
  Astrophys.\ J.\ Suppl.\  {\bf 148} (2003) 213
  [astro-ph/0302225].


\bibitem{hamann}
  J.~Hamann, L.~Covi, A.~Melchiorri and A.~Slosar,
  Phys.\ Rev.\ D {\bf 76} (2007) 023503
 [astro-ph/0701380];
\bibitem{covi}
  L.~Covi, J.~Hamann, A.~Melchiorri, A.~Slosar and I.~Sorbera,
  Phys.\ Rev.\ D {\bf 74} (2006) 083509
  [astro-ph/0606452].


\bibitem{Mortonson:2009qv} 
  M.~J.~Mortonson, C.~Dvorkin, H.~V.~Peiris and W.~Hu,
  Phys.\ Rev.\ D {\bf 79}, 103519 (2009)
  [arXiv:0903.4920 [astro-ph.CO]].

\bibitem{Hazra:2010ve} 
  D.~K.~Hazra, M.~Aich, R.~K.~Jain, L.~Sriramkumar and T.~Souradeep,
  JCAP {\bf 1010}, 008 (2010)
  [arXiv:1005.2175 [astro-ph.CO]].


\bibitem{Ashoorioon:2006wc} 
  A.~Ashoorioon and A.~Krause,
  hep-th/0607001.








\bibitem{Adshead:2012xz}
  P.~Adshead and W.~Hu,
  Phys.\ Rev.\ D {\bf 85} (2012) 103531
  [arXiv:1203.0012 [astro-ph.CO]].

\bibitem{Adshead:2011jq}
  P.~Adshead, C.~Dvorkin, W.~Hu and E.~A.~Lim,
  Phys.\ Rev.\ D {\bf 85} (2012) 023531
  [arXiv:1110.3050 [astro-ph.CO]].

\bibitem{Planck:2013kta} 
  P.~A.~R.~Ade {\it et al.}  [Planck Collaboration],
  arXiv:1303.5075 [astro-ph.CO].

\bibitem{Lewis:2002ah} 
  A.~Lewis and S.~Bridle,
  Phys.\ Rev.\ D {\bf 66}, 103511 (2002)
  [astro-ph/0205436].

\bibitem{camb}
  A.~Lewis, A.~Challinor and A.~Lasenby,
  Astrophys.\ J.\  {\bf 538}, 473 (2000)
  [arXiv:astro-ph/9911177].

\bibitem{Powell}
  M. J. D. Powell [ University of Cambridge Report],
 DAMTP 2009/NA06.

\bibitem{Ade:2013hta}
  P.~A.~R.~Ade {\it et al.}  [ Planck Collaboration],
  arXiv:1303.5072 [astro-ph.CO].

\bibitem{wmap9}
G.~Hinshaw, D.~Larson, E.~Komatsu, D.~N.~Spergel, C.~L.~Bennett, J.~Dunkley, M.~R.~Nolta and M.~Halpern {\it et al.},
arXiv:1212.5226 [astro-ph.CO].

\bibitem{meerburg}
 P.~D.~Meerburg and D.~N.~Spergel,
  arXiv:1308.3705 [astro-ph.CO].

\end{thebibliography}
\end{document}